\documentclass[12pt]{article}%
\usepackage{amssymb}
\usepackage{amsfonts}
\usepackage{sw20elba}%
\usepackage{amsmath}%
\usepackage{graphicx}
\providecommand{\U}[1]{\protect\rule{.1in}{.1in}}

\begin{document}

\title{Comparison of the pressure dependences of $T_{c}$ in the trivalent
$d$-electron superconductors \\Sc, Y, La, and Lu to Mbar pressures}
\author{M. Debessai, J. J. Hamlin, and J. S. Schilling\vspace{0.3cm}
\and \textit{Department of Physics, Washington University}\\\textit{CB 1105, One Brookings Dr, St. Louis, MO 63130}}
\date{June 17, 2008 }
\maketitle

\begin{abstract}
Whereas dhcp La superconducts at ambient pressure with $T_{c}\simeq5$ K, the
other trivalent $d$-electron metals Sc, Y, and Lu only superconduct if high
pressures are applied. Earlier measurements of the pressure dependence of
$T_{c}$ for Sc and Lu metal are here extended to much higher pressures.
Whereas $T_{c}$ for Lu increases monotonically with pressure to 12.4 K at 174
GPa (1.74 Mbar). $T_{c}$ for Sc reaches 19.6 K at 107 GPa, the 2nd highest
value observed for any elemental superconductor. At higher pressures a phase
transition occurs whereupon $T_{c}$ drops to 8.31 K at 111 GPa. The $T_{c}(P)$
dependences for Sc and Lu are compared to those of Y and La. An interesting
correlation is pointed out between the value of $T_{c}$ and the fractional
free volume available to the conduction electrons outside the ion cores, a
quantity which is directly related to the number of $d$ electrons in the
conduction band.

\end{abstract}

\newpage

\section{Introduction}

One of the most important goals in the field of superconductivity is to
recognize the properties favorable for pushing the superconducting transition
temperature $T_{c}$ to ever higher values. A material which superconducts at
or above room temperature would likely have a lasting impact on current
technology. The vast majority of known superconducting materials exhibit
$T_{c}$'s below 10 K, including all elemental superconductors at ambient
pressure \cite{kittel,schilling2}, as seen in Fig.~1. Values of $T_{c}$ at or
above 20 K have been found only for the cuprate \cite{cuprates} and Fe-based
oxides \cite{kamihara}, Nb$_{3}$Ge \cite{nbge}, MgB$_{2}$ \cite{nagamatsu},
Rb$_{3}$C$_{60}$ \cite{rbc60}$,$ possibly Cs$_{3}$C$_{60}$ \cite{palstra1},
and, under extreme pressures, for the elemental metals Ca \cite{yabuuchi} and
Y \cite{hamlin2}. Whereas the high-$T_{c}$ oxides, which superconduct at
temperatures as high as 134 K under ambient conditions (HgBa$_{2}$Ca$_{2}%
$Cu$_{3}$O$_{8+\delta}$) \cite{aschilling}, are generally believed to benefit
from a non-phononic pairing interaction \cite{cuprates}, the other members of
the above 20K-or-above group likely exhibit conventional electron-phonon
pairing. Whatever the nature of the pairing, it is important to establish the
conditions favorable for maximizing $T_{c}$.

High pressure is a particularly valuable tool for identifying systematics in a
given physical phenomenon, such as superconductivity or magnetism, since it
generates changes in the physical properties of a single sample in a
controlled manner. For example, in a simple-metal superconductor such as Al,
In, Sn, or Pb, where the conduction band is made up of $s,p$ electrons,
$T_{c}$ is always found to \textit{decrease} initially under pressure
\cite{schilling2}. In fact, this result was observed for Sn by Sizoo and Onnes
in 1925\ \cite{sizoo} in the first high-pressure experiment ever carried out
on a superconductor; they concluded that \textquotedblleft... a relatively
large space between the atoms is favourable for the appearing of the
supraconductive state ...\textquotedblright. It is, therefore, surprising that
$dT_{c}/dP$ is strongly positive for the simple metals Li
\cite{schillingalkali} and Ca \cite{yabuuchi} if they are subjected to
pressures in the range above 20 GPa. In superconductors containing transition
metals, where \textit{d}-electrons dominate the conduction band properties,
$T_{c}$ can initially rise or fall under pressure and exhibit a highly
nonlinear $T_{c}(P)$ dependence at higher pressures (see, for example, Refs.~
\cite{buzea,tissen1,shelton1,shelton2}). With such complexity in $T_{c}(P),$
it would seem useful to search for a simple, underlying mechanism to account
for the observed changes in $T_{c}$ as a function of decreasing interatomic
spacing as pressure is applied.

More than three decades ago Johansson and Rosengren \cite{johansson} pointed
out an interesting correlation between the crystal structure sequence
hcp$\rightarrow$Sm-type$\rightarrow$dhcp$\rightarrow$fcc across the rare-earth
series from Lu to La at ambient pressure, or for a given rare-earth metal
under increasing pressure, and the increasing fractional atomic volume
occupied by the ion core. This correlation was put on a more quantitative
footing by Merkau \textit{et al.} \cite{merkau} through their extensive
structural experiments to higher pressures and temperatures. Duthie and
Pettifor \cite{duthie1} showed that the observed structural sequence across
the rare-earth series both at ambient and high pressure can be quantitatively
correlated with the $d$-band occupancy $N_{d}.$ In fact, the crystal
structures across the 3$d$, 4$d$, and 5$d$ transition metal series have also
been shown to be closely correlated with $N_{d}$ \cite{Nd-structure reference}.

In view of this significant correlation between $N_{d}$ and crystal structure
for $d$-electron metals, it would be interesting to inquire whether other
properties, such as the value of the superconducting transition temperature
$T_{c},$ might also be correlated with the $d$-electron count $N_{d}$. As seen
in Fig.~1, with the exception of the magnetic (Cr, Mn, Fe, Co, Ni) and nearly
magnetic (Pd, Pt) transition metals, all transition-metal elements are
superconducting at ambient pressure with transition temperatures ranging from
325 $\mu$K for Rh to 9.50 K for Nb \cite{kittel,schilling2}. The fifteen
trivalent rare-earth metals La through Lu possess a similar $d$-electron
character near the Fermi energy as the beginning transition metals Sc and Y,
neither of which is superconducting at ambient pressure. Of the rare earths,
only La superconducts at ambient pressure, the remaining, besides Yb and Lu,
being magnetic which acts to suppress the superconductivity. We note that an
interesting systematics in $T_{c}$ was uncovered by McMillan \cite{mcmillan}
for the 5$d$-electron transition metal series; the empirical electronic
density of states at the Fermi energy $N(E_{f})$ estimated from $T_{c}$ and
specific heat data was found to track well a calculated canonical electronic
density of states dependence.

In this paper we examine whether there is a correlation between the change in
$T_{c}$ under pressure and the $d$-electron count $N_{d}$ which increases
under pressure as the fractional atomic volume of the ion core increases. As
the relative increase in $N_{d}$ with pressure is particularly large for the
\textquotedblleft early\textquotedblright\ transition metals where $N_{d}$ is
small, we focus our attention first on the four trivalent $d$-electron metals
Sc, Y, La, and Lu, of which only La superconducts at ambient pressure.
Previous high-pressure studies on Sc \cite{hamlin1}, Y \cite{hamlin2}, La
\cite{tissen1}, and Lu \cite{wittig1} were restricted to pressures of 74, 115,
50, and 28 GPa, respectively. Here we present new data which determine
$T_{c}(P)$ for Sc and Lu to the significantly higher pressures of 123 and 174
GPa, respectively. Whereas in Lu $\ T_{c}$ increases monotonically with
pressure to 12.4 K at 174 GPa, in Sc $\ T_{c}$ increases rapidly with
pressure, reaching a maximum value of 19.6 K at 107 GPa in the Sc-II phase. If
the pressure is increased further, Sc-II transforms to Sc-III
\cite{fujihisa1,akahama1} whereupon $T_{c}$ drops to 8.31 K at 111 GPa. Sc
possesses with 19.6 K the second highest value of $T_{c}$ of any elemental
superconductor. An interesting correlation is revealed between the value of
$T_{c}$ for these four metals and the increasing fractional ion core volume
under pressure.

\section{Experiment}

The diamond anvil cell (DAC) used in the present experiment is made of both
standard and binary CuBe alloy and utilizes a He-gas-driven double-membrane to
change the force between the two opposing diamond anvils at any temperature
\cite{schilling21}. Temperatures as low as 1.55 K can be reached in an Oxford
flow cryostat. The 1/6-carat, type Ia diamond anvils have 0.18 mm culets
beveled at 7$^{\circ}$ out to 0.35 mm with a 3 mm girdle. The metal gasket is
a disc made of W$_{0.75}$Re$_{0.25}$ alloy 3 mm in diameter, 250 $\mu$m thick,
and preindented to 25-30 $\mu$m; a 90 $\mu$m diameter hole is electro-spark
drilled through the center of the gasket. High-purity ingots of Sc and Lu
(99.98\% metals basis) were obtained from the Materials Preparation Center of
the Ames Laboratory \cite{ames1}. Small chips of Sc or Lu were cut from the
ingots are then packed as densely as possible into the gasket hole. Several
tiny ruby spheres \cite{chervin} were placed next to the sample to allow the
determination of the pressure \textit{in situ} at 20 K from the R$_{1}$ ruby
fluorescence line with resolution $\pm$ 0.2 GPa using the revised pressure
scale of Chijioke \textit{et al.} \cite{silvera1}. An Ar-ion or HeCd laser was
used to excite the ruby fluorescence. To maximize the sample diameter under
extreme pressure conditions, and thus the magnitude of the diamagnetic signal
at the superconducting transition, no pressure medium was used in the present
experiments. In previous experiments on Y \cite{hamlin2}, no measureable
difference was observed in the pressure dependence of $T_{c}$ with (dense He)
or without pressure medium in the 35 - 90 GPa pressure region where they could
be compared. One should not forget, however, that in nonhydrostatic
experiments employing no pressure medium, shear stress effects may have a
significant influence on how $T_{c}$ changes under pressure
\cite{schilling2,deemyad1,stishov1}.

The highest pressure reached in the present experiments was 174 GPa (1.74
Mbar) for Lu. As can be seen in Fig.~2a, this extreme pressure is sufficient
to cause the nominally flat culet surface of the diamond anvils to cup
which\ leads to the black halo around the Lu sample in reflected white-light
illumination. At this 174 GPa pressure the ruby line became too weak to be
detected. In this case the pressure was determined from the first order Raman
spectrum \cite{akahama} of the diamond anvil (see Fig.~2b) taken from a
spherical region $\sim$ 20 $\mu$m in diameter centered on the Lu sample (blue
region in Fig.~2a). The Raman signal from outside of this region was rejected
by the confocal microscope optics.

The superconducting transition is detected inductively using two compensating
primary/secondary coil systems (see Fig.~2c) connected to a Stanford Research
SR830 digital lock-in amplifier via a SR554 transformer preamplifier; the
excitation field for the ac susceptibility studies is 3 Oe r.m.s. at 1023 Hz.
Under these conditions and considering the calibration of the coil system, the
anticipated diamagnetic signal in nanovolt for a superconducting transition
with 100\% shielding is given from Ref.~ \cite{hamlinthesis} by%
\begin{equation}
S(nV)=8.17\times10^{-5}[V\backslash(1-\mathcal{D})],\label{signal}%
\end{equation}
where $V$ is the sample volume in ($\mu$m)$^{3}$ and $\mathcal{D}$ is the
demagnetization factor. Since the sample is a flat cylinder, $V=\pi hd^{2}/4,
$ where $h$ and $d$ are the sample thickness and diameter, respectively. In
the limit $h/d<<1,$ Joseph \cite{joseph} has derived the approximate
expression
\begin{equation}
\mathcal{D}\approx1-\left[  2h/(\pi d)\right]  \left[  \ln\left(  8d/h\right)
-1\right]  .\label{joseph}%
\end{equation}
In the present experiments to extreme pressure the sample dimensions are
typically $d\simeq$ 80 $\mu$m and $h\simeq$ 15 $\mu$m, yielding $\mathcal{D}%
\approx0.671$ and thus $S\approx20$ nV. A more accurate calculation
\cite{chen} finds $\mathcal{D}\approx0.73$ and thus $S\approx25 $ nV.

To facilitate the identification of the superconducting transition, a
temperature-dependent background signal $\chi_{b}^{\prime}\left(  T\right)  $
is subtracted from the measured susceptibility data; $\chi_{b}^{\prime}\left(
T\right)  $ is obtained by measuring at pressures too low to induce
superconductivity over the temperature range 5-50 K. This lower limit in the
effective temperature range is not dictated by the cryostat, which can cool to
1.55 K, but by the fact that the superconductivity of the W-Re gasket leads to
a large diamagnetic signal which swamps the sample signal below $\sim$ 5.2 K,
a temperature which has negligible pressure dependence. For this reason the
superconducting transition of the sample can only be reliably detected if it
occurs at a temperature $T_{c}\gtrsim5.5$ K. A relatively low noise level of a
few tenths of a nanovolt is achieved by: (a) using the transformer
preamplifier to ensure good impedance matching, (b) varying the temperature
very slowly (100~mK/min) at low temperatures, (c) using a long time constant (%
$>$
3 s) on the lock-in amplifier, and (d) averaging over 2-3 measurements.
Further experimental details of the high pressure and ac susceptibility
techniques are published elsewhere \cite{hamlin1,schilling21,deemyad1}.

\section{Results of Experiment}

\subsection{Sc Metal}

The initial pressurization of the Sc sample was carried out at room
temperature. The force between the diamond anvils was gradually increased
until the gasket hole completely closed around the sample, compressing it to
full density. At this point the pressure on the sample was approximately 20
GPa and the sample diameter had decreased from 90 to 85 $\mu$m. Increasing the
pressure to 35 GPa resulted in no further decrease in the sample diameter. The
height of the hole in the gasket containing the sample varied between the
initial preindentation thickness of 26 $\mu$m and the final thickness after
the experiment 17 $\mu$m; we estimate the sample thickness during the
high-pressure experiment to be 17-20 $\mu$m.

Following the initial pressurization to 35 GPa, the DAC was cooled to low
temperatures to search for a superconducting transition. None was observed
above 5.5 K in the ac susceptibility at 35, 56, or 66 GPa, whereby the DAC was
kept at a temperature below 160 K to expedite the experimentation. After
warming the DAC back to ambient temperature, 81 GPa pressure was applied and
the DAC cooled down and kept below 160 K for the rest of the experiment. As
seen in Fig.~3a, at 81 GPa a superconducting transition does appear\ where
$T_{c}$ increases with pressure to a value as high as 19.6 K at 107 GPa, but
then drops to a much lower temperature at 123 GPa. The magnitude of the
superconducting transition ($\sim$ 20-30 nV), which is consistent with 100\%
shielding, is much larger than that ($\sim$ 3-4 nV)\ in the previous nearly
hydrostatic experiments on Sc by Hamlin \textit{et al.} \cite{hamlin1} to 74.2
GPa. This is due to the larger sample volume and larger demagnetization factor
in the present nonhydrostatic experiments. Note that we define $T_{c}$ as the
temperature at the midpoint of the diamagnetic transition.

In Fig.~3b the dependence of $T_{c}$ for Sc on pressure is plotted for all
data in the present experiment (unprimed numbers) and compared to the earlier
high-pressure studies of Wittig \cite{wittig2} to 21.5 GPa in solid steatite
pressure medium (solid line) and those of Hamlin \textit{et al.}
\cite{hamlin1} to 74.2 GPa in nearly hydrostatic He pressure medium (primed
numbers). For all three sets of data using diverse pressure media, the
dependence of $T_{c}$ on pressure appears to follow a reasonably smooth,
monotonically increasing curve to 107 GPa. As we also found for Y
\cite{hamlin2}, therefore, the $T_{c}(P)$-dependence for Sc does not appear to
depend sensitively on the degree of shear stress on the sample. We note,
however, that the absence of a superconducting transition above 5.5 K in the
present experiment at 66 GPa does appear to conflict with the nearly
hydrostatic data point \#2$^{\prime}$ of Hamlin \textit{et al.} \cite{hamlin1}
in Fig.~3b where $T_{c}\approx6.2$ K at 66.8 GPa, thus pointing to possible
minor shear stress effects on $T_{c}(P)$ in the present experiment.

Between 0 and 123 GPa, the highest pressure reached in the present
experiments, Sc undergoes two structural phase transitions (see phase
boundaries in Fig.~3b) \cite{fujihisa1,akahama1}. Whereas no $T_{c}(P)$ data
is available across the I$\rightarrow$II boundary, $T_{c}$ is seen to drop
sharply at the II$\rightarrow$III boundary and then rise slowly as the
pressure is increased further. The value of $T_{c}\simeq$ 19.6 K
(susceptibility midpoint) reached at 107 GPa shortly before the II$\rightarrow
$III phase transition is the second highest value of $T_{c}$ ever observed in
an elemental superconductor, trailing only Ca with $T_{c}\approx$ 25 K
(resistivity onset) at 160 GPa \cite{yabuuchi}. Note that the highest value
reached for Y is $T_{c}\simeq$ 19.5 K (susceptibility midpoint) at 115 GPa
\cite{hamlin2}.

As expected for a superconducting transition, $T_{c}$ decreases in a dc
magnetic field. The transition in Fig.~3a at 102 GPa was measured after a 500
Oe magnetic field was applied at 25 K (solid red line). The dependence of
$T_{c}$ at this pressure on magnetic field $H$ to 500 Oe is shown in the inset
to Fig.~3a and is seen to decrease approximately linearly with $H$ at the rate
$dT_{c}/dH\simeq-0.30$ mK/Oe. For the superconducting transitions in Sc at 81,
87, 97, 102, 111, and 123 GPa, where $T_{c}(H=0)\equiv T_{co}$ = 10.6, 12.8,
17.4, 18.9, 8.31, and 8.85 K, $dT_{c}/dH$ takes on the values -0.63, -0.56,
-0.49, -0.30, -0.78, and -0.78 mK/Oe, respectively. Hamlin \textit{et al
}\cite{hamlin1}\textit{\ }reported for data point 3$^{\prime}$ in Fig.~3b,
where $T_{c}\approx8.2$ K, that $\left\vert dT_{c}/dH\right\vert \leq0.3$
mK/Oe\textit{. }For an Y sample with $T_{c}\simeq9.7$ K at 46.6 GPa, $T_{c}$
was found to decrease under magnetic fields to 500 Oe at the rate -0.5 mK/Oe,
a comparable value to those found for Sc \cite{hamlin2}.

An attempt to extend the present experiment on Sc to pressures above 123 GPa
resulted in the destruction of one of the two diamond anvils, thus ending the experiment.

\subsection{Lu Metal}

A single high pressure ac susceptibility experiment was carried out on pure Lu
metal. The W$_{0.75}$Re$_{0.25}$ gasket was preindented to 29 $\mu$m and, as
for Sc, the Lu sample was loaded into the 90 $\mu$m diameter gasket hole. The
DAC was then pressurized at ambient temperature to $\sim$ 20 GPa whereupon the
diameter of the bore containing the sample decreased from 90 to 83 $\mu$m. The
DAC was then cooled to low temperatures to search for a superconducting
transition in the temperature range above 5.5 K, a limit dictated, as before
with Sc, by the superconducting transition of the W-Re gasket below 5.2 K. No
superconducting transition was detected above 5.5 K at pressures of 40, 62,
and 69 GPa. Finally, at 88 GPa a strong diamagnetic transition was observed
near 7 K, as seen in Fig.~4a. Inserting the observed sample diameter of $\sim$
80 $\mu$m and estimated 15 $\mu$m$\ $thickness into Eq.(\ref{signal}), a value
for the diamagnetic signal $S\approx$ 20 nV is obtained. Since the measured
transitions in Fig.~4a lie near $30$ nV, the indicated diamagnetic shielding
is at or near 100\%. Given the tiny sample size, the quality of the data is
quite remarkable.

$T_{c}$ for Lu was found to increase monotonically with pressure to 140 GPa,
at which point the He-gas pressure $P_{mem}$ in the double-membrane reached 45
bars. At higher pressures we could no longer detect the ruby R$_{1}$ line. The
superconducting transition of Lu was measured to higher pressures by
increasing $P_{mem}$ from 45 to 80 bars. For $P_{mem}\leq$ 45 bars, the
dependence of the sample pressure (from the ruby R$_{1}$ line) on $P_{mem}$ is
well described by a simple linear fit, making a linear extrapolation of this
curve to higher pressures seem reasonable. Such an extrapolation yields an
estimated sample pressure of 220 GPa for $P_{mem}\approx$ 80 bar. To check the
validity of this extrapolation, we measured the diamond vibron in the Raman
scattering (Fig.~2b) at the maximum pressure, as described above, which
yielded `only' 174 GPa. The simple extrapolation from $P_{mem}\approx$ 45 to
80 bar thus overestimated the sample pressure in the cell by more than 40 GPa.
This reduction in the actual pressure likely arises at least in part from
progressive \textquotedblleft cupping\textquotedblright\ of the diamond anvil
culet surface at extreme pressures, as evidence by the black annular region
clearly visible in Fig.~2b.

In Fig.~4b $T_{c}$ for Lu is plotted versus pressure for all data in the
present experiment. Four of the transitions (points 6-9) occur beween 140 and
174 GPa where we made no direct measurement of the pressure. For these points
(open circles) we estimate the sample pressure from $P_{mem}$ using a linear
interpolation between 140 GPa at 45 bars and 174 GPa at 80 bars. That the
dependence of $T_{c}$ on pressure for Lu is reversible is evidenced by the
fact that data point 11, obtained by releasing the pressure from 174 to 120
GPa, lies along the $T_{c}(P)$-curve for increasing pressure.

Lu has been found to transform at room temperature from a dhcp to hR24
structure near 88 GPa and remain in this structure up to at least 163 GPa
\cite{chesnut}. Unfortunately, our data do not extend to low enough pressure
to allow us to comment on the possible effect of this structural transition on
$T_{c}(P).$ In this experiment there was no catastrophic failure of the
diamond anvils upon complete release of pressure. One of the two anvils did
show a ring-crack pattern typical for beveled anvil experiments in this
pressure range \cite{eremetsbook}.

The dependence of $T_{c}$ on dc magnetic fields $H$ up to 500 Oe was measured
at most of the pressures. In Fig.~4c we show superconducting transitions for
Lu measured at 140 GPa and fields of 0, 167, 333, and 500 Oe. The transition
temperature decreases monotonically and reversibly with $H $, as expected for
a superconductor. No difference in behavior was observed whether the dc field
was applied above or below $T_{c}.$ The initial slope $dT_{c}/dH\approx-0.6$
mK/Oe remains constant over the entire pressure range studied. Unlike Sc, the
magnitude of the diamagnetic transition for Lu is seen to become\ noticeably
smaller with increasing field. This likely arises since the applied field,
which is enhanced by the factor $(1-\mathcal{D})^{-1}$ at the outer perimeter
of the pancake-shaped sample, is sufficiently strong for $T<T_{c}$ to exceed
the critical field and penetrate into the outer perimeter of the sample, thus
reducing the effective sample diameter $d $ and sample volume $V.$ In fact,
from the relative change in the magnitude of the superconducting transition in
the applied magnetic field seen in Fig.~4c, one can estimate \cite{note30} the
critical field at 0 K and 140 GPa for Lu to be $H_{c}$(0 K$)\approx1440$ Oe.

In contrast, as seen in Fig.~3a for Sc at 102 GPa, there is no measurable
decrease in the magnitude of $S$ in 500 Oe magnetic field. The enhanced
magnetic field is thus too small to penetrate into the perimeter of the
disk-shaped sample. We can, therefore, only put a lower limit on the size of
the critical field $H_{c}$(0 K$)\gtrsim H_{o}(1-\mathcal{D})^{-1}.$ Since from
Eq.(\ref{joseph}) for $h\simeq17$ $\mu$m and $d\simeq85$ $\mu$m (see above) it
follows that $\mathcal{D}\simeq0.658,$ for Sc at 102 GPa we estimate that
$H_{c}$(0 K$)\gtrsim(500$ Oe$)(1-0.658)^{-1}=$ 1460 Oe.

\section{Discussion}

\subsection{Phenomenological Model}

In Fig.~5a we directly compare the pressure dependences of $T_{c}$ for Sc and
Lu with the results of previous studies on the other trivalent $d$-electron
metals Y \cite{hamlin2,note1} and La \cite{tissen1}. $T_{c}(P)$ for Y and La
appears to pass through a maximum value at $\sim$ 120 and 12 GPa,
respectively, however, $T_{c}(P)$ for La displays considerably more structure
over its pressure range to 50 GPa than for the other three to over 100 GPa.
This may be at least partly a result of the relatively high compressibility of
La metal. In Fig.~5b we utilize the measured equations of state of Sc
\cite{fujihisa1}, Y \cite{grosshans5}, La \cite{grosshans5}, and Lu
\cite{chesnut}\textit{\ }to convert the data in Fig.~5a to plots of $T_{c} $
versus relative volume $V/V_{o},$ where $V_{o}$\ is the sample volume at
ambient pressure \cite{note7}. Note that for Sc, Y, and Lu the dependence of
$T_{c}$ on $V/V_{o}$ exhibits a positive curvature over an appreciable region.

We first attempt a simple phenomenological analysis of the volume dependences
of $T_{c}$ in Fig.~5b using the McMillan equation%
\begin{equation}
T_{c}\simeq\frac{\left\langle \omega\right\rangle }{1.20}\exp\left[
\frac{-1.04\left(  1+\lambda\right)  }{\lambda-\mu^{\ast}\left(
1+0.62\lambda\right)  }\right]  ,\label{mcmillan}%
\end{equation}
where $\lambda\equiv\eta/(M\left\langle \omega^{2}\right\rangle )$ is the
electron-phonon coupling parameter, $\eta$ the Hopfield parameter, $\omega$ a
phonon frequency, $\mu^{\ast}$ the Coulomb repulsion, and $M$ the ionic mass
\cite{mcmillan}. If we define $\gamma\equiv-\partial\ln\left\langle
\omega\right\rangle /\partial\ln V$ and $\varphi\equiv\partial\ln
\lambda/\partial\ln V$, assume $\gamma$ and $\varphi$ are independent of
pressure, and integrate, we obtain%
\begin{equation}
\left\langle \omega\right\rangle _{V}=\left\langle \omega\right\rangle _{o}
\left[  V/V_{o}\right]  ^{-\gamma}\text{ \ \ \ \ and \ \ \ \ \ }%
\lambda(V)=\lambda(V_{o})\left[  V/V_{o}\right]  ^{\varphi},\label{vol dep}%
\end{equation}
where $\varphi\simeq\partial\ln\eta/\partial\ln V+2\gamma.$ The parameter
$\partial\ln\eta/\partial\ln V$ is negative and normally lies near -1 for
$s,p$-metals or -3 to -5 for $d$-metals \cite{schilling2}. Since $2\gamma$ is
positive, whether $\lambda$ (and $T_{c}$) increases or decreases with pressure
depends on whether $\left\vert d\ln\eta/d\ln V\right\vert >2\gamma$ or vice
versa. The next step is to fix the values of $\left\langle \omega\right\rangle
_{o}$ and $\gamma$ from experimental data \cite{parameters1}, set $\mu^{\ast
}=0.1$, and then find the best fit to the dependence of $T_{c}$ on relative
volume in Fig.~5b by using $\lambda(V_{o})$ and $\partial\ln\eta/\partial\ln
V$ as fit parameters \cite{hamlinthesis}. Since at ambient pressure dhcp La
superconducts with $T_{c}(V_{o})\simeq$ 5 K \cite{tissen1},
Eq.~(\ref{mcmillan}) is used to determine $\lambda(V_{o})$ for La.

As seen in Fig.~5b, the fits obtained using Eq.~(\ref{mcmillan}) are
reasonably successful. For Y there is a clear change in slope near
$V/V_{o}\approx0.63$ where a structural phase transition occurs (Sm-type
$\rightarrow$ dhcp) \cite{grosshans5,vohra5} so that two fits are carried out,
one for the \textquotedblleft low-$T_{c}$\textquotedblright\ and the other for
the \textquotedblleft high-$T_{c}$\textquotedblright\ values. The values of
the two fit parameters used $\left(  \lambda(V_{o}),\partial\ln\eta
/\partial\ln V\right)  $ are found to be (0.166, -4.15), (0.127, -5.50)
(0.459, -2.83), (0.844,-4.03), and (0.366, -2.81) for Sc, Y(low-$T_{c}$),
Y(high-$T_{c}$), La, and Lu, respectively, allowing the estimate that at
ambient pressure $T_{c}$ equals 0 K, 0 K, 1.2 K, 5.0 K, and 0.31 K. From
experiment it is known that $T_{c}(V_{o})<6$ mK for Y \cite{wittig1},
$<$
30 mK for Sc \cite{ScTc}, and $<22$ mK for Lu \cite{note3} which agrees
reasonably well with the estimates above. Note that from the Y(high-$T_{c}$)
fit $T_{c}(V_{o})\approx1.2$ K is predicted, meaning that if Y would remain
metastable in its dhcp phase at ambient pressure, it should superconduct at
$T_{c}\approx1.2$ K, a value more than two orders of magnitude higher than
that ($<6$ mK) in its thermodynamically stable hcp structure. Extrapolating
the fit curves in Fig.~5b to higher pressures leads to the estimate that,
barring structural transitions, $T_{c}$ would reach 30 K at 127, 164, 53, and
580 GPa for Sc, Y(high-$T_{c}$), La, and Lu, respectively.

\subsection{Electronic Structure Calculations}

The above phenomenological analysis shows that the $T_{c}(P)$ dependences
observed for Sc, Y, La, and Lu appear consistent with moderately
strong-coupled, phonon-mediated superconductivity using reasonable values of
the averaged parameters. However, to pinpoint the mechanism(s) responsible for
the significant increase in $T_{c}$ with pressure in experiment, detailed
electronic structure calculations are needed. Nixon \textit{et al.}
\cite{nixon1} recently used an augmented plane wave (APW) method to calculate
the electronic structure of Sc assuming, for simplicity, an fcc phase. Over
the pressure range 20 to 80 GPa they find that the Hopfield parameter $\eta$
increases by nearly a factor of four, whereas the electronic density of states
$N(E_{f})$ decreases by 15\%, the Coulomb repulsion $\mu^{\ast}$ decreasing by
only 5\%. Using the McMillan formula in Eq.~(\ref{mcmillan}), they find that
over the given pressure range to 80 GPa $T_{c}$ increases from 0.4 to 7 K, in
reasonable agreement with experiment.

The same group \cite{lei1} used similar techniques to estimate the electronic
structure of fcc Y to pressures somewhat above 1 Mbar (113 GPa). They estimate
that over the pressure range 40 to 113 GPa $T_{c}$ increases by 5 - 10 K,
depending on the value chosen for $\mu^{\ast},$ the best agreement occurring
for $\mu^{\ast}=0.04.$ Linear response methods were applied by Yin \textit{et
al.} \cite{yin6} who included pressure-dependent changes in the lattice
vibration spectrum of fcc Y metal in their calculation. They conclude that the
large positive value of $dT_{c}/dP$ arises from a pressure-induced softening
in the transverse phonon modes, i.e. a negative mode Gr\"{u}neisen parameter,
in contrast to the positive value $\gamma\simeq1.08$ used in the above
phenomenological analysis. Singh \cite{singh} has recently applied
density-functional theory to both hcp and dhcp Y metal to calculate the
changes under pressure of both the electronic properties and the lattice
vibration spectrum. A substantial increase in the electron-phonon coupling
with pressure is found yielding a value of $T_{c}$ for dhcp Y as high as 19 K.

Some time ago Pickett \textit{et al.} \cite{pickett1} carried out a linearized
APW calculation for fcc La to 12 GPa. They find that, as with Sc and Y, the
strong increase of $T_{c}$ with pressure arises primarily from a significant
enhancement of the Hopfield parameter $\eta.$ In their DAC studies on La to 50
GPa, Tissen \textit{et al.} \cite{tissen1} suggest that the abrupt increase in
$T_{c}$ near 2 GPa likely arises from the dhcp$\rightarrow$fcc structural
phase transition, whereas some of the marked features in $T_{c}(P)$ at higher
pressures may arise because of $s\rightarrow d$ transfer which pushes the
Fermi energy up through van Hove singularities.

In 1990 Skriver and Mertig \cite{skriver8} calculated the strength of the
electron-phonon coupling parameter $\lambda$ at ambient pressure, obtaining
for Sc, Y, La, and Lu the values 0.57, 0.53, 0.90, and 0.59, respectively.
Note that the only ambient-pressure superconductor in the group, La, has a
much higher calculated value of $\lambda$ than the other three.

It would be useful if a single state-of-the-art electronic structure
calculation of the properties relevant for superconductivity would be carried
out for Sc, Y, La, and Lu to pressures into the Mbar region. Because of the
close electronic similarity of these four systems, much could be learned about
the efficacy of this type of calculation for predicting superconducting
properties in general.

\subsection{d-Band Occupancy}

As mentioned in the Introduction, the equilibrium crystal structure under
ambient conditions across the 3$d$, 4$d$, and 5$d$ transition metal series, as
well as across the rare-earth series from La to Lu, has been shown to be
closely related to the occupancy of the $d$-band $N_{d}.$ We now explore the
question whether in $d$-electron metals the superconducting transition
temperature $T_{c}$ might itself be correlated with $N_{d},$ restricting
ourselves here to the four electronically closely related trivalent
$d$-electron metals Sc, Y, La, and Lu.

The $d$-electron count $N_{d}$ increases under pressure due to $s\rightarrow
d$ transfer which is driven by the increase in the fractional ion core volume
$V_{c}/V_{a}$ \cite{duthie1,pettifor1}, where we define the ion core volume
$V_{c}\equiv(4/3)\pi R_{c}^{3}$ and the atomic volume $V_{a}\equiv(4/3)\pi
R_{WS}^{3}$ ($R_{WS}$ is the Wigner-Seitz radius), yielding $R_{WS}%
/R_{c}=(V_{a}/V_{c})^{1/3}.$ The conduction electrons must stay out of the ion
core volume $V_{c}$ and thus are confined to the free sample volume
$V_{f}\equiv V_{a}-V_{c}$ outside the ion cores. Under pressure the atomic
volume $V_{a}$ decreases whereas $V_{c}$ remains nearly constant. The ratio
$R_{WS}/R_{c},$ therefore, is a measure of how much free volume remains for
the conduction electrons under pressure. The ratio $R_{WS}/R_{c}$ decreases
under pressure; the closer it approaches the minimum possible value 1, the
less free volume is available and the greater the anticipated degree of $s-d$
transfer \cite{duthie1,pettifor1}.

Many years ago Johansson and Rosengren \cite{johansson} showed that the
$T_{c}$ values for Y, La, Lu, and alloys thereof are a smooth function of a
similar ratio \cite{note2} which decreases under pressure, as does $T_{c}$. We
pursue a similar analysis here where $R_{WS}$ at ambient pressure is
calculated from the molar volume and $R_{c}$ is obtained from the trivalent
ionic radii for coordination number 6 \cite{springer1}. We assume that $R_{c}
$ is independent of pressure so that applying high pressure monotonically
\textit{decreases} the value of the ratio $R_{WS}/R_{c}$. To determine how
$R_{WS}/R_{c}$ changes at high pressure, we simply multiply it by $\left(
V/V_{o}\right)  ^{1/3},$ where $V/V_{o}$ is given by the equations of state
for Sc, Y, La, and Lu cited above.

In Fig.~6 we plot $T_{c}$ versus $R_{WS}/R_{c}$ for Sc, Y, La, and Lu. One
sees immediately that the data for these four metals are more tightly grouped
together than in the previous figures where $T_{c}$ was plotted versus
pressure $P$ or relative volume $V/V_{o}.$ The ratio $R_{WS}/R_{c}$,
therefore, appears to be a more relevant parameter to describe the
superconducting properties than $P$ or $V/V_{o}.$ Some simple systematics are
evident in Fig.~6. Initially, at least, $T_{c}$ generally increases with
pressure. Interestingly, the $T_{c}$ values of all four elements do not exceed
1 K until the ratio $R_{WS}/R_{c}$ is reduced to a value below $\sim$ 2.1.
This clarifies why La is the only member in this group that is superconducting
at ambient pressure; for La at ambient pressure $R_{WS}/R_{c}=2.02$, whereas
for the other three metals $R_{WS}/R_{c}>2.1$ (see vertical arrows below upper
axis in Fig.~6).

Under pressure the $d$-electron count $N_{d}$ for Sc, Y, La, and Lu increases
\cite{duthie1,pettifor1,pickett3}. Duthie and Pettifor \cite{duthie1} and
Pettifor \cite{pettifor1} have shown that the occupation of the $d$-band is
closely related to the fractional volume of the ion core with smaller relative
volumes leading to greater occupation of the $d$-band. The effect of
compression on the $d$-band occupancy has been recently calculated for Sc, Y,
La, and Lu by Yin and Pickett \cite{pickett3} and is shown in Fig.~7a where
$N_{d}$ is plotted versus $V/V_{o}$. Note that $N_{d}$ increases monotonically
with pressure (decreasing $V/V_{o})$, being largest for La metal over almost
the entire range. Fig~7b shows that for Sc, Y, and La the ratio $R_{WS}/R_{c}$
has nearly a one-to-one correspondence with the calculated $d$-electron count
$N_{d}$, the dependence for Lu being shifted towards lesser $N_{d}$ values.

In Fig.~8 the data in Figs.~5b and 7a are used to plot $T_{c}$ versus $N_{d}$
for all four metals. Compared to the data in Fig.~6, where $T_{c}$ is plotted
versus the ratio $R_{WS}/R_{c}$, the curves for Y, La, and Lu do appear to be
grouped closer together, but that for Sc has moved somewhat further away. It
is thus not clear whether the ratio $R_{WS}/R_{c}$ or the $d $-electron count
$N_{d}$ is the superior parameter for describing changes in the superconducing
properties under pressure.

Since $T_{c}$ generally increases with $N_{d},$ one expects that when the $d$
occupation reaches its maximum value $N_{d}=3,$ the pressure dependence of
$T_{c}$ should change, perhaps passing through a maximum. According to
Fig.~7b, however, the principal maximum in $T_{c}(P)$ for La occurs at a value
$N_{d}\approx2.4$ which is well below $N_{d}=3$. Sc, being the least
compressible and having the largest ambient pressure value of $R_{WS}/R_{c},$
is the farthest from completion of $s-d$ transfer in the present experiment.
Indeed, the data for Sc in Fig.~7b would imply that it would take a pressure
much higher than 2 Mbar before $s-d$ transfer is completed. This suggests
that, had the structural phase transition in Sc at 110 GPa not occurred,
$T_{c}$ might have reached values near 30 K according to an estimate using the
phenomenological model above.

The Sc-II phase, in which Sc exhibits its highest value $T_{c}\simeq19.6$ K,
the second highest behind Ca \cite{yabuuchi} for any elemental superconductor,
is an unusual incommensurate host-guest crystal structure \cite{fujihisa1}.
This type of crystal structure was only recently found to exist in high
pressure phases of elemental solids \cite{degtyareva}. It would be very
interesting to study these metals to much higher pressures in order to
investigate to what heights $T_{c}$ for Sc and Lu will increase. In view of
its light molecular weight and exceptionally high value of $T_{c}$, ultra-high
pressure experiments on Sc are particularly promising. In addition, Sc
undergoes a further structural transition to Sc-IV at 130 GPa \cite{fujihisa1}
which may well leave its mark on the superconducting properties.

\vspace{0.4cm}

\noindent Acknowledgements. \ The authors would like to express their
gratitude to Z. Yin and W. E. Pickett for providing the data in figure 7a
before publication. The assistance of B. Wopenka in taking the Raman spectrum
of the diamond vibron. Thanks are due R. W. McCallum and K. W. Dennis of the
Materials preparation Center, Ames Lab, for providing the high purity Sc and
Lu samples. The authors are grateful to V. K. Vohra for recommending the
specifications for the beveled diamond anvils used in these experiments. The
authors also acknowledge research support by the National Science Foundation
through grant DMR-0703896.

\begin{center}
{\LARGE Figure Captions}

\bigskip

\bigskip
\end{center}

\noindent\textbf{Fig.~1.} \ Periodic Table listing 30 elements which
superconduct at ambient pressure (yellow) and 22 elements which only
superconduct under high pressure (green). For each element the upper position
gives the value of $T_{c}$(K) at ambient pressure; middle position gives
maximum value $T_{c}^{\max}$(K) reached in a high-pressure experiment at
$P$(GPa) (lower position). In many elements multiple phase transitions occur
under pressure. If $T_{c}$ decreases under pressure, only the ambient pressure
value of $T_{c}$ is given. Except for Sc and Lu, sources for $T_{c}$ values at
ambient and high pressure are given in Ref.~\cite{schilling2}.\bigskip

\noindent\textbf{Fig.~2.} \ (a) Micrograph in reflected white light of Lu
sample (blue) in W-Re gasket at 174 GPa; black annular ring signals cupping of
the diamond culet (180 $\mu$m diameter) at these extreme pressures. (b) Raman
spectrum from center of diamond anvil culet. High-energy edge of diamond
vibron spectrum at 1650 cm$^{-1}$ corresponds to pressure of 174 GPa
\cite{akahama}. (c) Two identical compensating primary/secondary coils systems
(each 180 turns of 60 $\mu$m diameter Cu wire) for ac susceptibility
measurements. Active coil is around 16-facet diamond anvil in middle;
compensating coil contains a W-Re dummy gasket. \bigskip

\noindent\textbf{Fig.~3.} \ (a) Real part of the ac susceptibility signal in
nanovolts (nV) versus temperature for Sc at selected pressures to 123 GPa.
Pressure was increased monotonically. Applying 500 Oe dc magnetic field shifts
superconducting transition at 102 GPa to lower temperatures (red curve).
Inset: \ $T_{c}$ versus magnetic field $H$ at 102 GPa. Vertical bars give
error in shift of $T_{c}$ using the transition for $H=0$ as reference. (b)
Superconducting transition temperature versus pressure in present experiment
($\blacksquare,$ unprimed numbers), from Ref.~\cite{hamlin1} ($\bullet,$
primed numbers), and from Ref.~\cite{wittig2} (short solid line).
\textquotedblleft Error bars\textquotedblright\ give 20-80 transition width.
Numbers give order of measurements. Dashed line through data is guide to eye.
Vertical dashed lines mark phase boundaries I$\rightarrow$II and
II$\rightarrow$III.\bigskip

\noindent\textbf{Fig.~4.} \ (a) Real part of the ac susceptibility signal in
nanovolts (nV) versus temperature for Lu at 88, 140, and 174 GPa pressure. (b)
Dependence of $T_{c}$ on pressure for all data. \textquotedblleft Error
bars\textquotedblright\ give 20-80 transition width. Numbers give order of
measurement. Dashed line through data is guide to eye. At 75 GPa (point 12) no
superconducting onset was observed above 5.2 K. Filled circles ($\bullet)$
indicate pressure measured from ruby R$_{1}$ line, open circles ($\circ)$
indicate pressure estimated from double-membrane pressure (see text). (c)
\ Real part of the ac susceptibility signal versus temperature at 140 GPa for
dc magnetic fields 0, 167, 333, and 500 Oe.\bigskip

\noindent\textbf{Fig.~5.} \ (a) $T_{c}$ versus pressure for the trivalent $d
$-electron metals Sc ($\bullet$), Y ($\blacklozenge$), La (solid line), and Lu
($\blacksquare$). Dashed lines are guides to the eye. The pressure for the
open diamond ($\diamond$) data point for Y is extrapolated \cite{note1}. (b)
$T_{c}$ versus relative volume using the $T_{c}(P)$ data from Fig.~5a. Solid
lines are fits to the data using the McMillan equation (see text). For La only
the data for $V/V_{o}>0.92$\ are fit.\bigskip

\noindent\textbf{Fig.~6.} \ $T_{c}$ versus ratio of Wigner-Seitz radius to
core-electron radius $R_{WS}/R_{c}$ for Sc, Y, La, and Lu. Dashed lines are
guides to the eye. Vertical arrows at the upper axis show ambient pressure
values of $R_{WS}/R_{c}$\ for the indicated elements.\bigskip

\noindent\textbf{Fig.~7.} \ (a) Calculated occupation of $d$-band $N_{d}$
versus relative volume $V/V_{o}$ for Sc, Y, La, and Lu from Yin and Pickett
\cite{pickett3}. (b) $N_{d}$ versus the ratio $R_{WS}/R_{c}$ from the data in
(a). Solid lines connect calculated data points.\bigskip

\noindent\textbf{Fig.~8.} \ $T_{c}$ versus $N_{d}$ using data from Figs.~5b
and 7a. Vertical arrows at the upper axis show ambient pressure values of
$N_{d}$\ for the indicated elements.


\begin{thebibliography}{99}
\bibitem {note30}Both the measured change in the ac susceptibility signal at
the superconducting transition $\Delta S$ and the applied dc field $H_{o}=500
$ Oe at the sample perimeter are enhanced by the same factor $(1-\mathcal{D}%
)^{-1}$, where $\mathcal{D}$\ is the demagnetization factor. Thus the change
in the diamagnetic signal at zero applied dc field is given by $\Delta
S_{1}=AV_{1}\Delta\chi V_{1}(1-\mathcal{D}_{1})^{-1}$, where $\mathcal{D}%
_{1}=0.671$ (see text), and at $H_{o}=500$ Oe field $\Delta S_{2}=AV_{2}%
\Delta\chi(1-\mathcal{D}_{2})^{-1}.$ Here, $\Delta\chi=-1$ is the
susceptibility jump at the superconducting transition, $A$ is a calibration
constant, $V_{1}=h\pi d_{1}^{2}/4$ and $V_{2}=h\pi d_{2}^{2}/4$ are the sample
volumes at 0 and 500 Oe field, respectively$,$ where $h\simeq15$ $\mu$m is the
sample thickness and $d_{1}\simeq80$ $\mu$m and $d_{2}$ the sample diameters
(note that $d_{2}<d_{1}$ if the applied field penetrates into the sample
perimeter). From the above it follows that $(1-\mathcal{D}_{2})^{-1}%
=(d_{1}/d_{2})^{2}(\Delta S_{2}/\Delta S_{1})(1-\mathcal{D}_{1})^{-1};$ the
only unknown in this equation is $d_{2}$ after Eq.(\ref{joseph}) is
substituted for $\mathcal{D}_{2}$. From the data in Fig~4c, $\Delta
S_{2}/\Delta S_{1}\simeq$ (27.4 nV)/(34.3 nV) = 0.799. Solving the above
transcendental equation for $d_{2}$, one obtains $d_{2}\simeq73.5$ $\mu$m and
thus, from Eq.(\ref{joseph}), $\mathcal{D}_{2}\simeq0.653.$ The critical
magnetic field at 0 K is then given by $H_{c}($0 K$)=H_{o}(1-\mathcal{D}%
_{2})^{-1}\simeq$ (500 Oe)/(1 - 0.653) = 1440 Oe.
\bibitem {note1}The data point ($\diamond$) for Y at 125 GPa in Fig.~5a was
added to the published data ($\blacklozenge$) to 115 GPa from Ref.
\cite{hamlin2}. At this highest pressure the ruby manometer failed (intensity
too low); the sample pressure of 125 GPa was estimated by extrapolation from
the He-gas pressure $P_{mem}$ in the double-membrane, where $P_{mem}$ is
proportional to the force pushing the diamond anvils together.
\bibitem {note7}As discussed in Ref.~\cite{grosshans5}, the Murnaghan equation
\cite{murnaghan} can be written in the form $P=P_{r}+(B_{r}/B_{r}^{\prime
})[(V_{r}/V)^{B_{r}^{\prime}}-1]$ to fit partial $V(P)$ data which begins at a
reference pressure $P_{r}$, as for Sc from Ref.~ \cite{fujihisa1}. Here it is
used to fit $V(P)$ data on Sc to 104 GPa \cite{fujihisa1}, Y to 50 GPa
\cite{grosshans5}, and La to 40 GPa \cite{grosshans5}. Note that for an
experiment initiated at ambient pressure, $P_{r}=0.0001$ GPa $\approx$ 0 GPa,
$B_{r}=B_{o},$ $B_{r}^{\prime}=B_{o}^{\prime},$ and $V_{r}=V_{o}.$ This
Murnaghan equation is used to extrapolate $V(P)$ to the higher pressures
reached in the data shown in Fig.~5a (Sc to 123 GPa, Y to 117 GPa, and La to
50 GPa). The following values of $[P_{r}($GPa$),B_{r}($GPa$),B_{r}^{\prime}]$
give the best fits: \ Sc to 22 GPa (0, 78, 1.8), Sc 22 to 104 GPa (22, 140,
1.5) where a 7.0\% volume collapse occurs at 22 GPa \cite{fujihisa1}; Y to 50
GPa (0, 40, 2.4); La to 40 GPa (0, 23, 2.9). For Lu Chesnut and Vohra
\cite{chesnut} use the modified universal equation of state to fit their
structural data both in the pressure range to 88 GPa, where a 4.9\% volume
collapse occurs after the phase transition, and 88 to 163 GPa. We use this
equation to extrapolate $V(P)$ to the highest pressure in our experiment 174 GPa.
\bibitem {parameters1}For Sc, Y, La, and Lu values of the Gr\"{u}neisen
parameter $\gamma$ (Ref.~\cite{gschneidner1}) and Debye temperature
$\Theta_{D}$ used (Ref.~\cite{tsang}) are 1.10, 1.08, 0.80, 1.00 and 345 K,
244 K, 139 K, and 183 K, respectively, where$\left\langle \omega\right\rangle
\simeq0.69\Theta_{D}$ from Ref.~\cite{butler}.
\bibitem {note3}In Ref.~\cite{wittig1} Probst and Wittig state that Lu does
not superconduct above 22 mK at 0.8 GPa. They extrapolate their high-pressure
data to obtain the estimate $T_{c}\approx1$ mK at ambient pressure.
\bibitem {note2}In Ref. \cite{johansson} $R_{WS}$ is the Wigner-Seitz radius
which minimizes the total energy for a given vale of the pseudopotential core
radius $R_{c}.$ The basic idea is the same, but in the present reatment we use
empirically determined values for both radii.
\end{thebibliography}
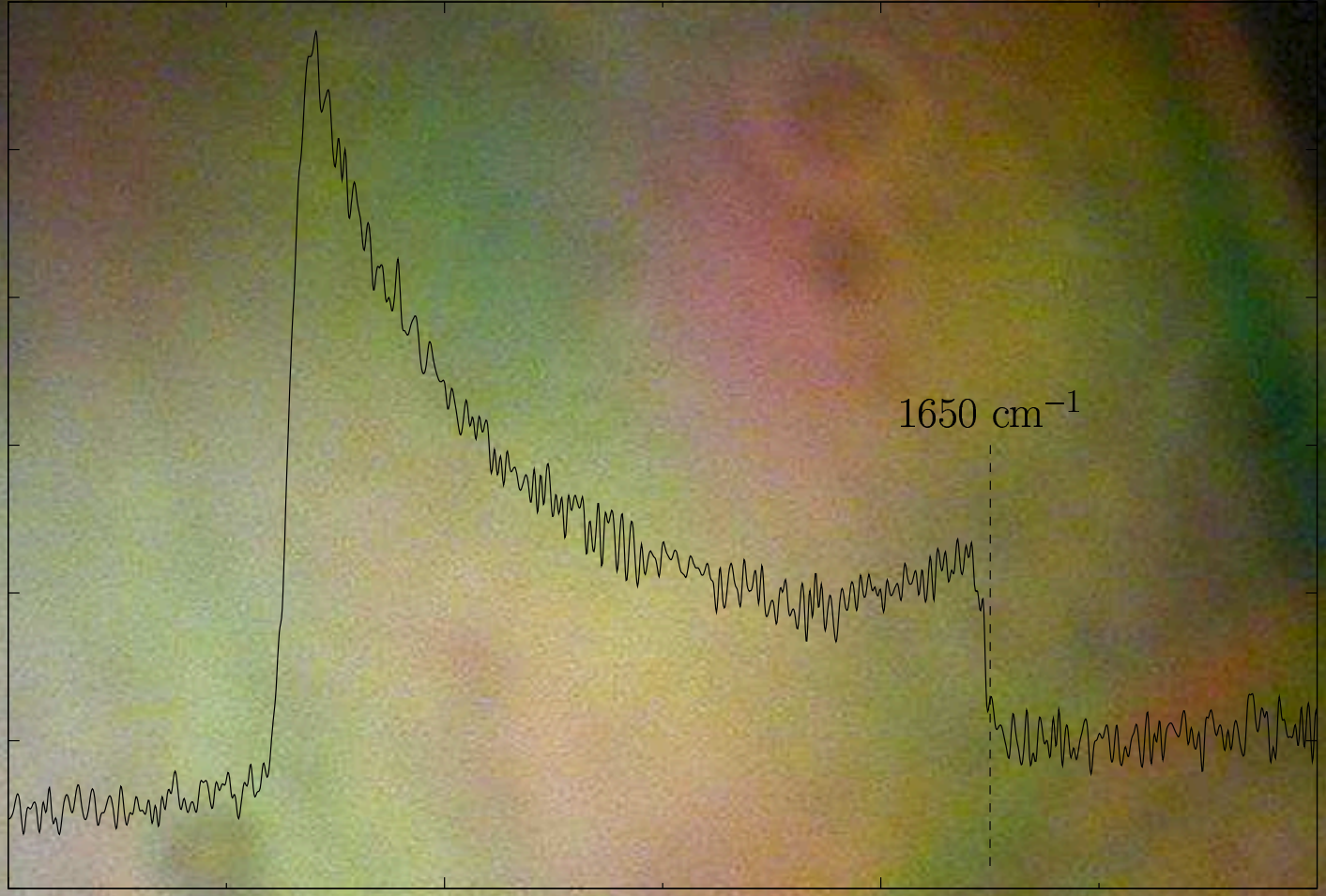
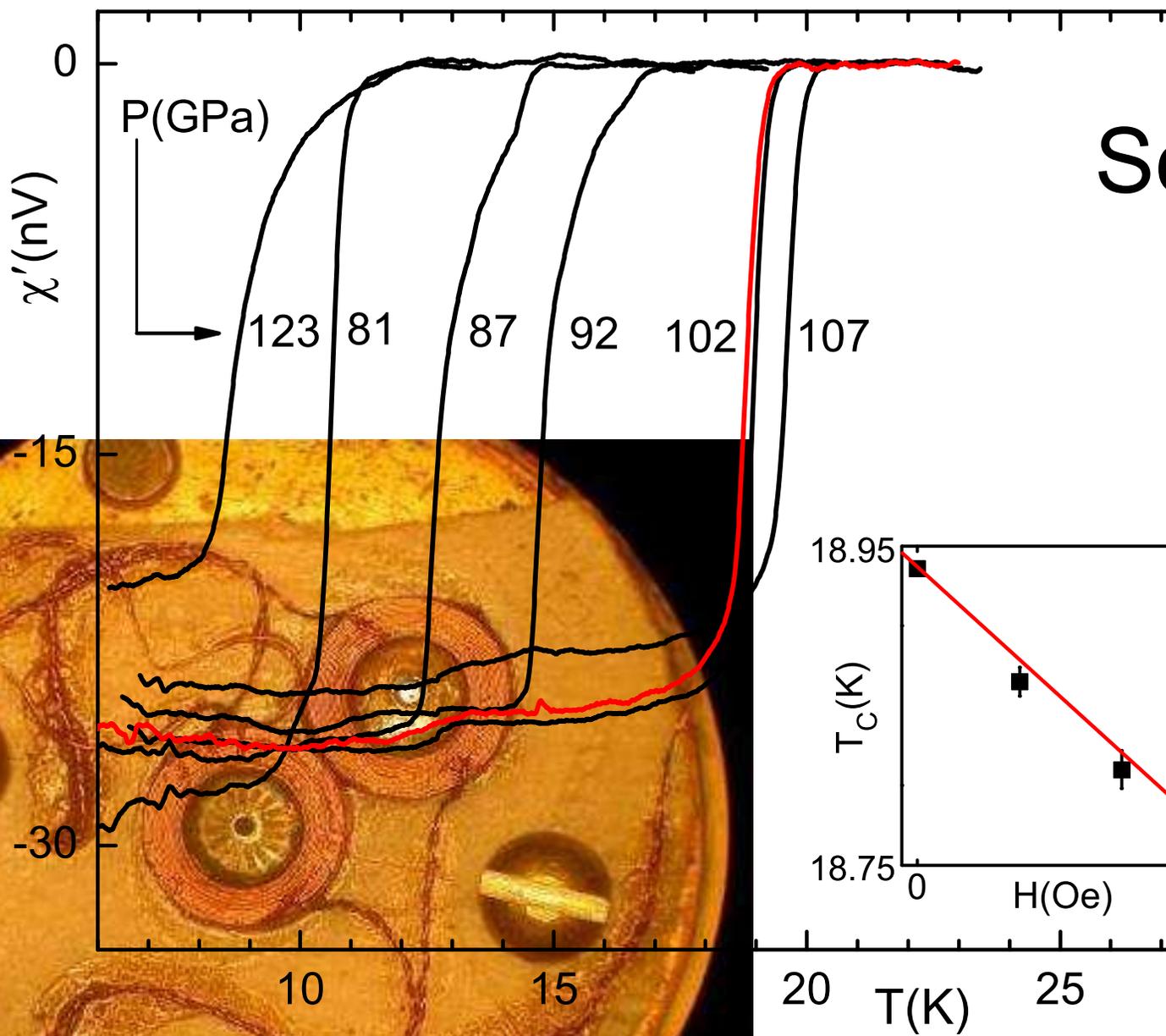
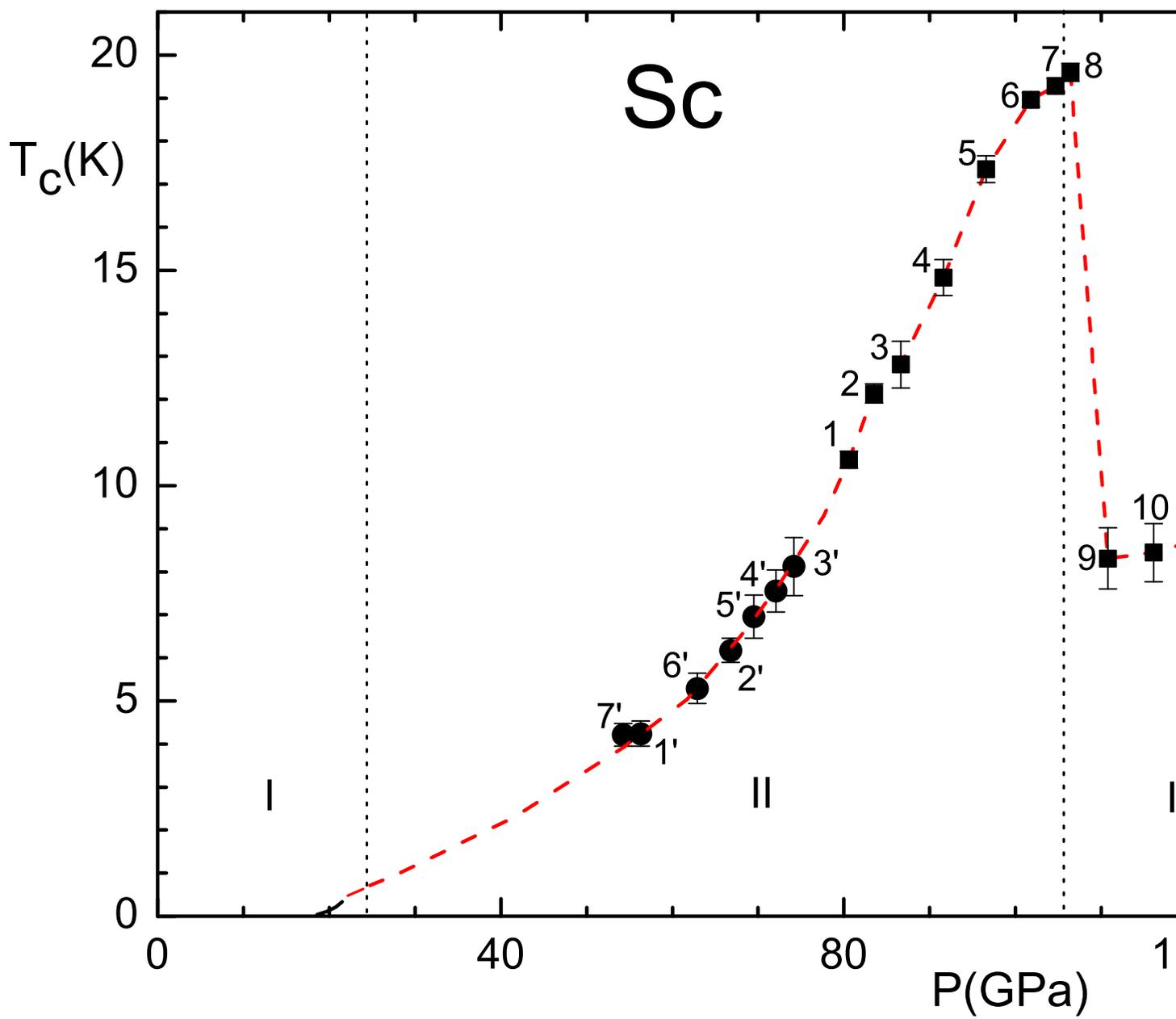
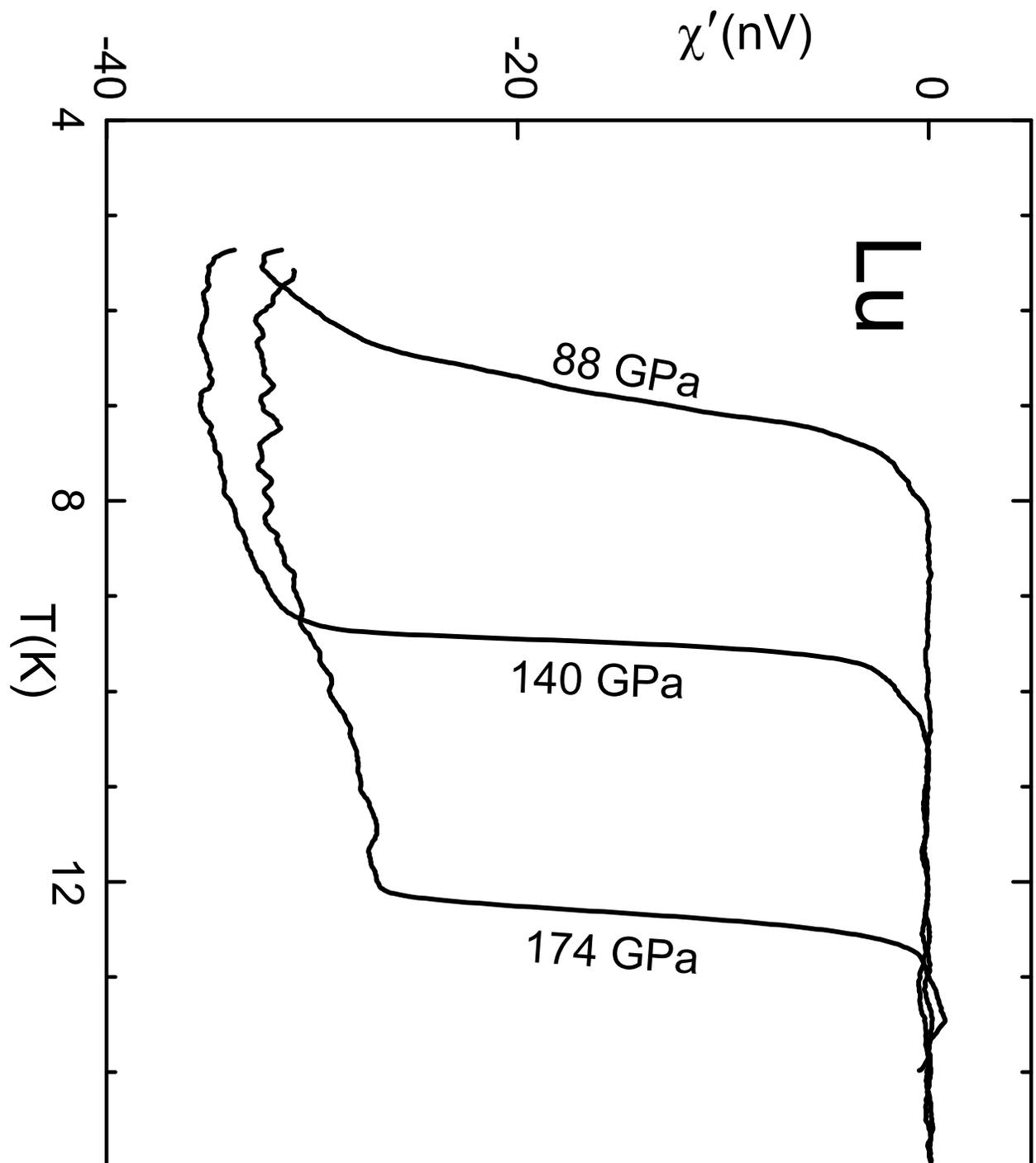
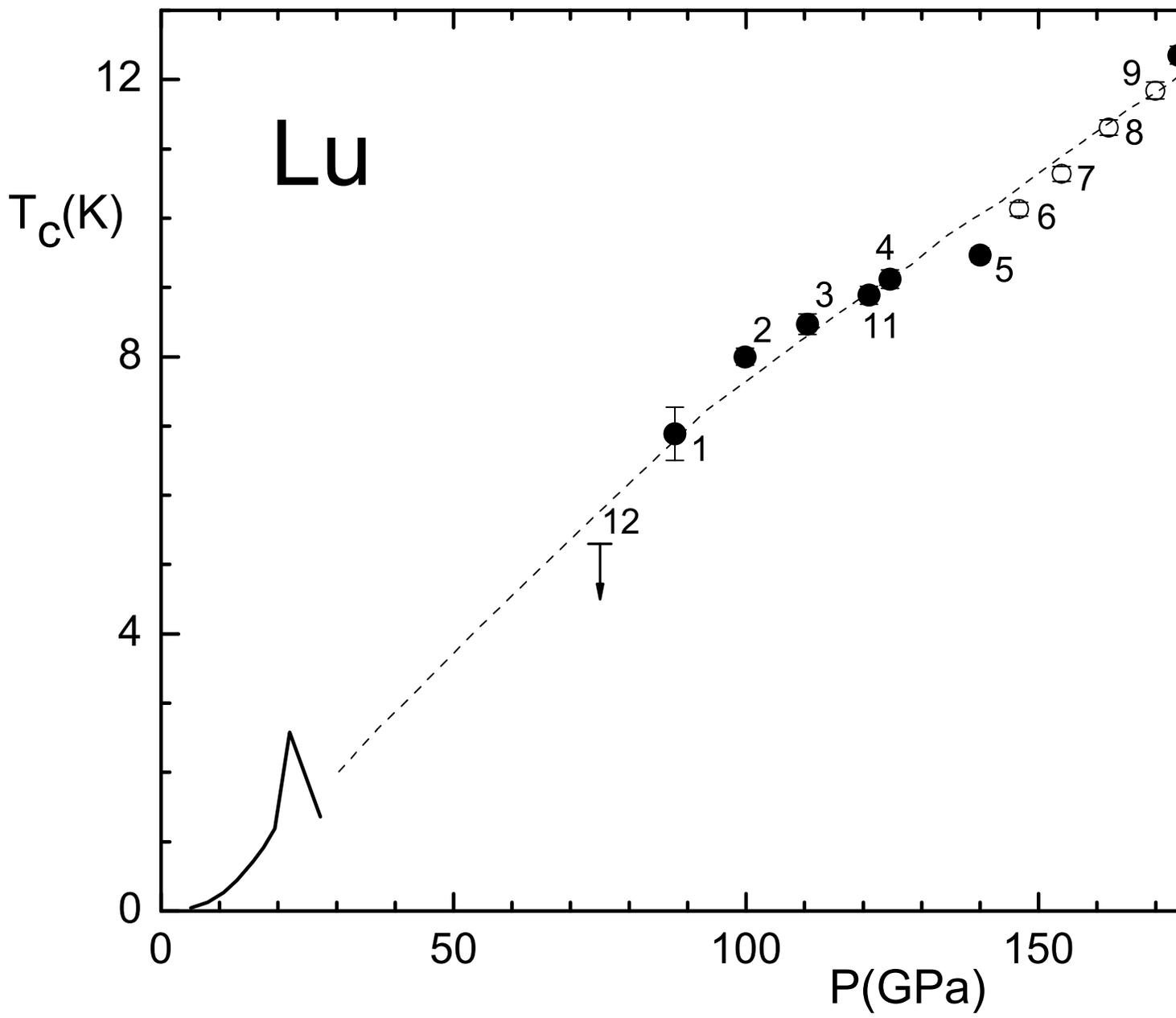
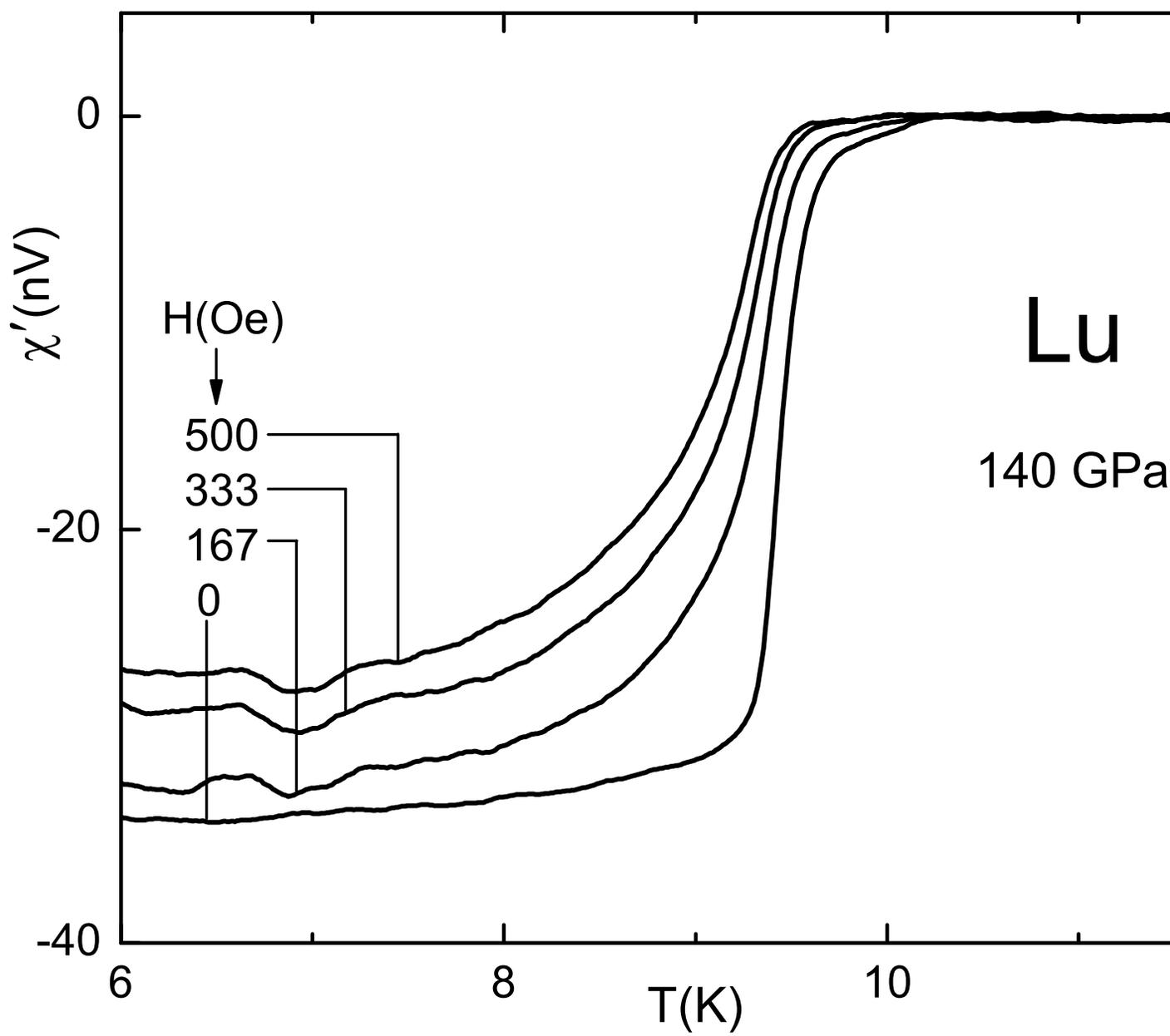
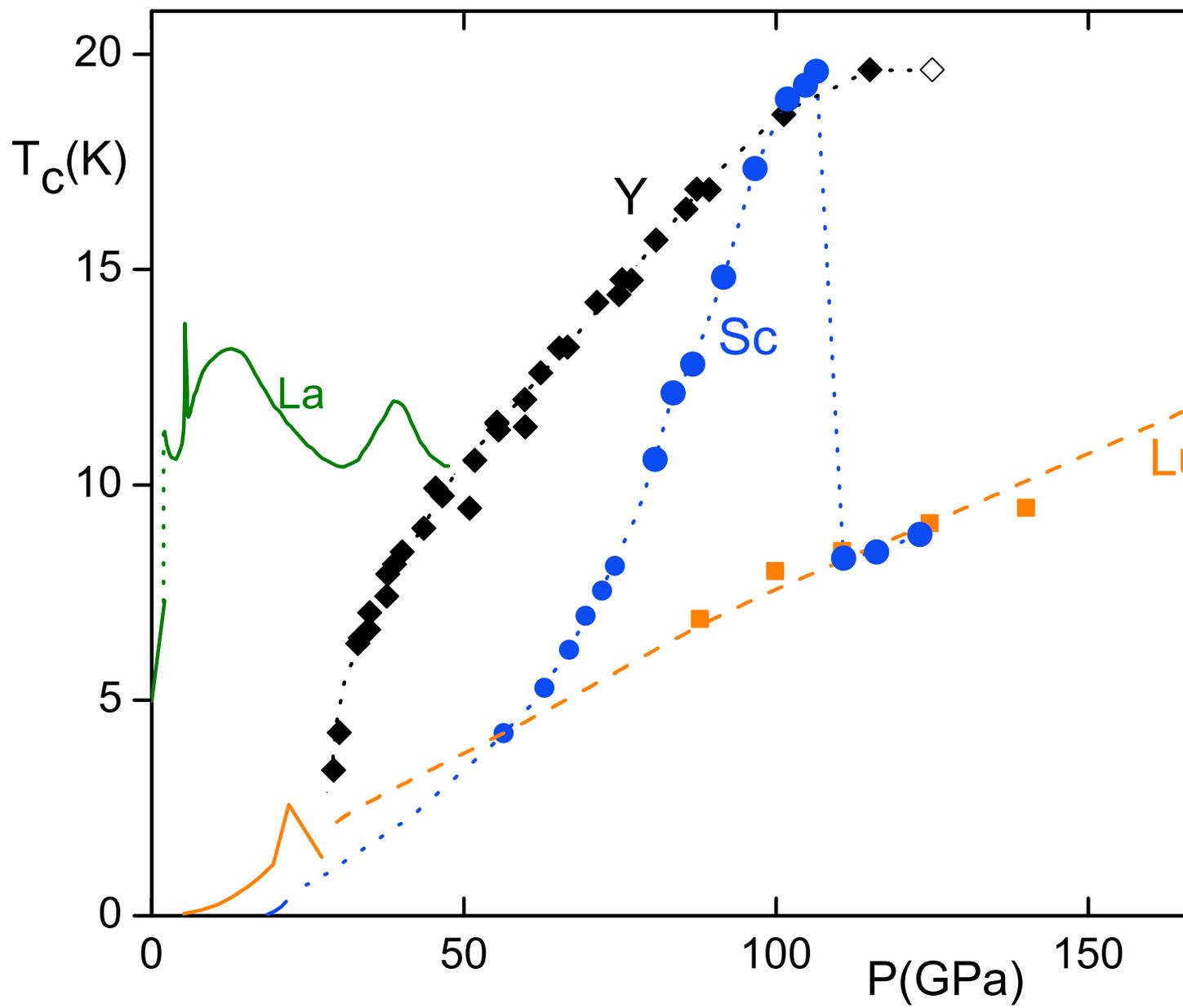
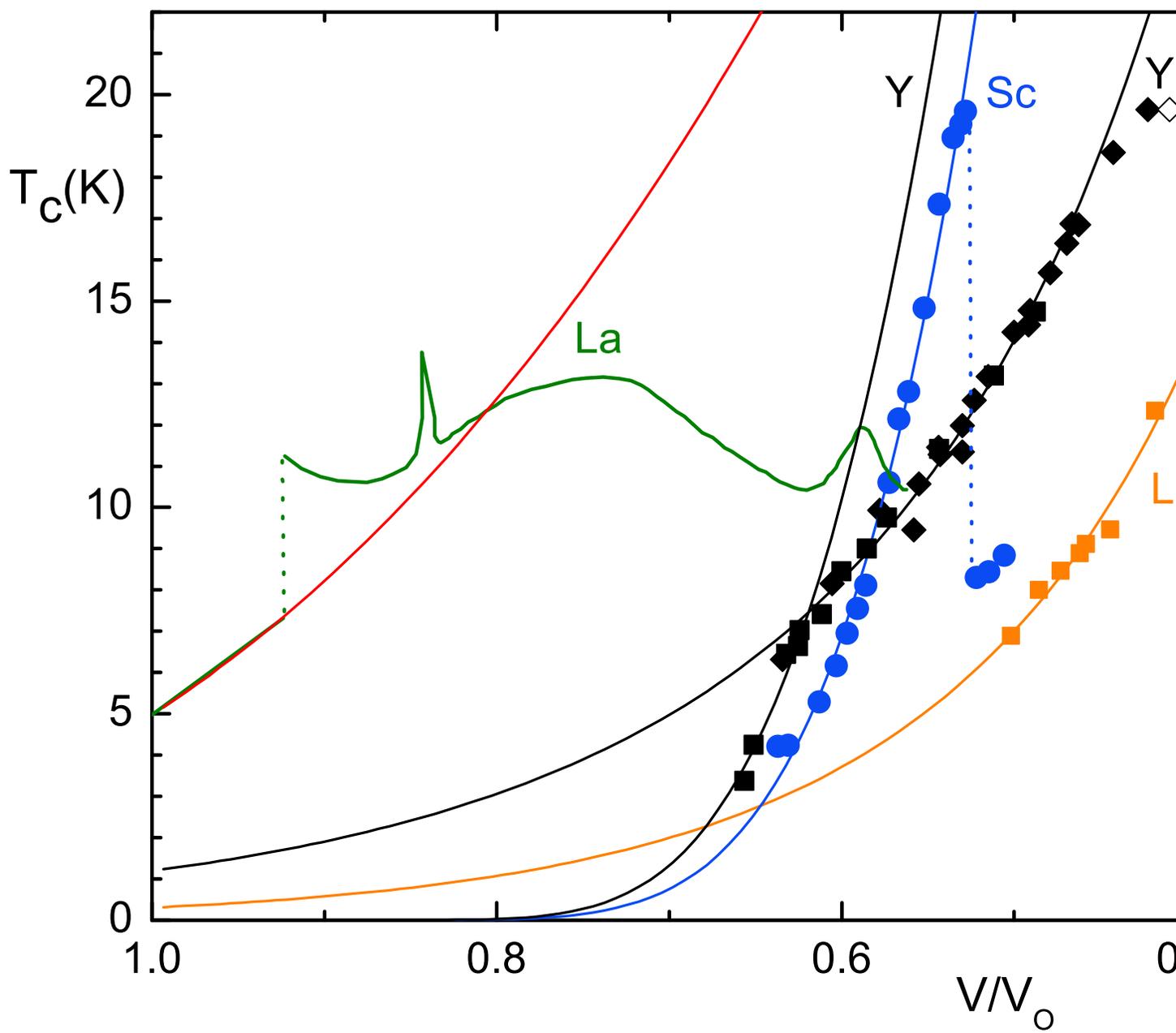
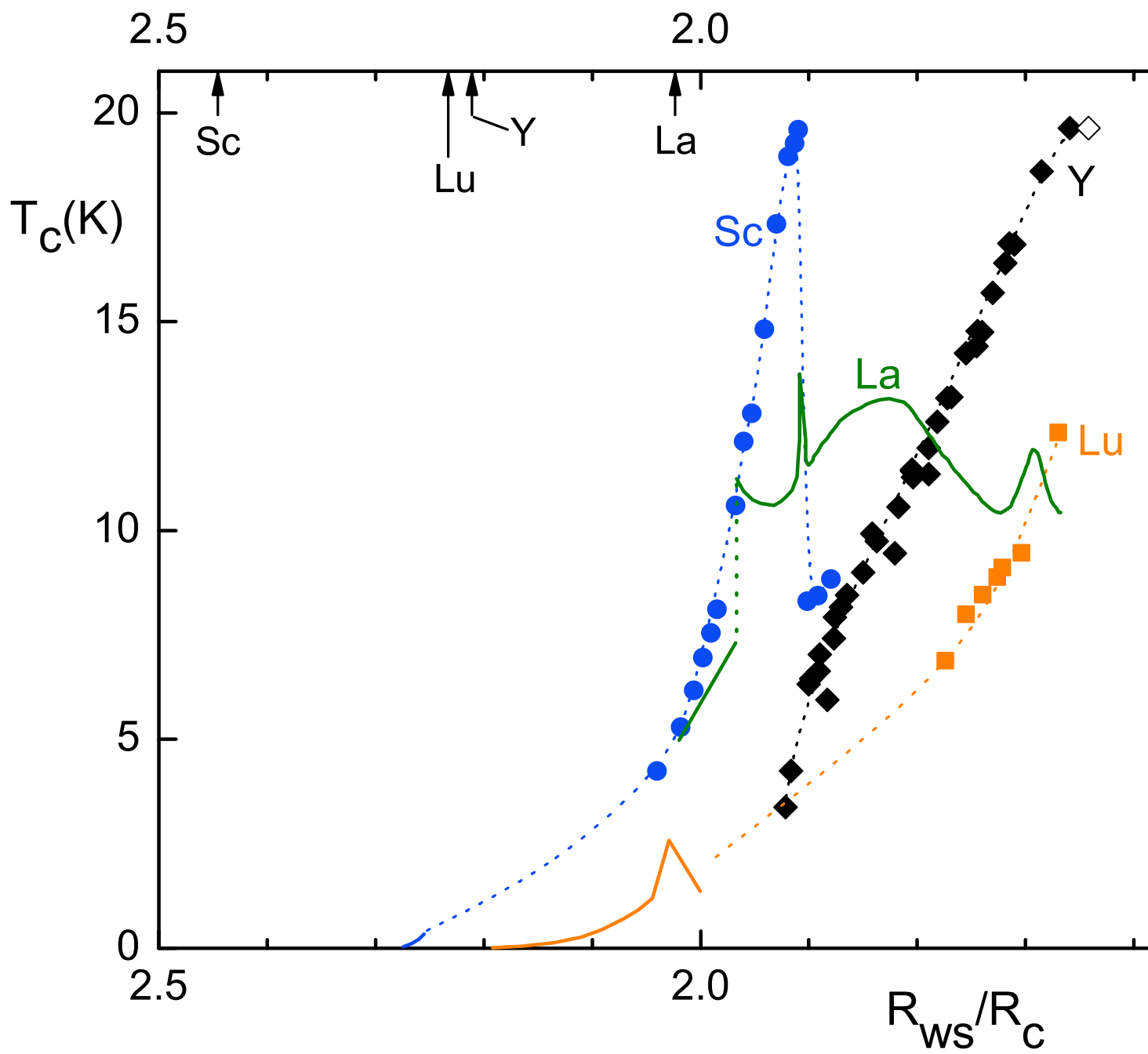
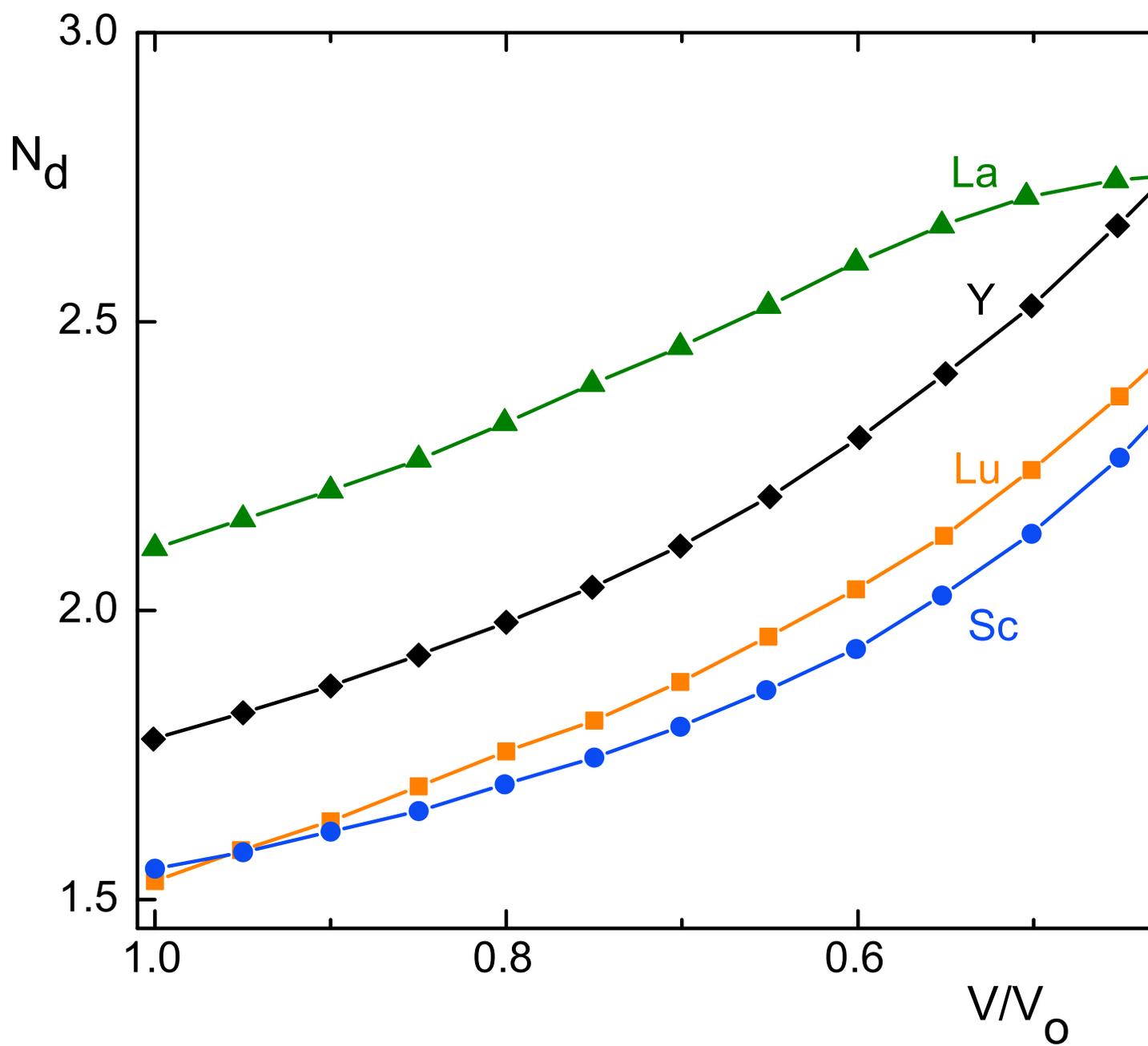
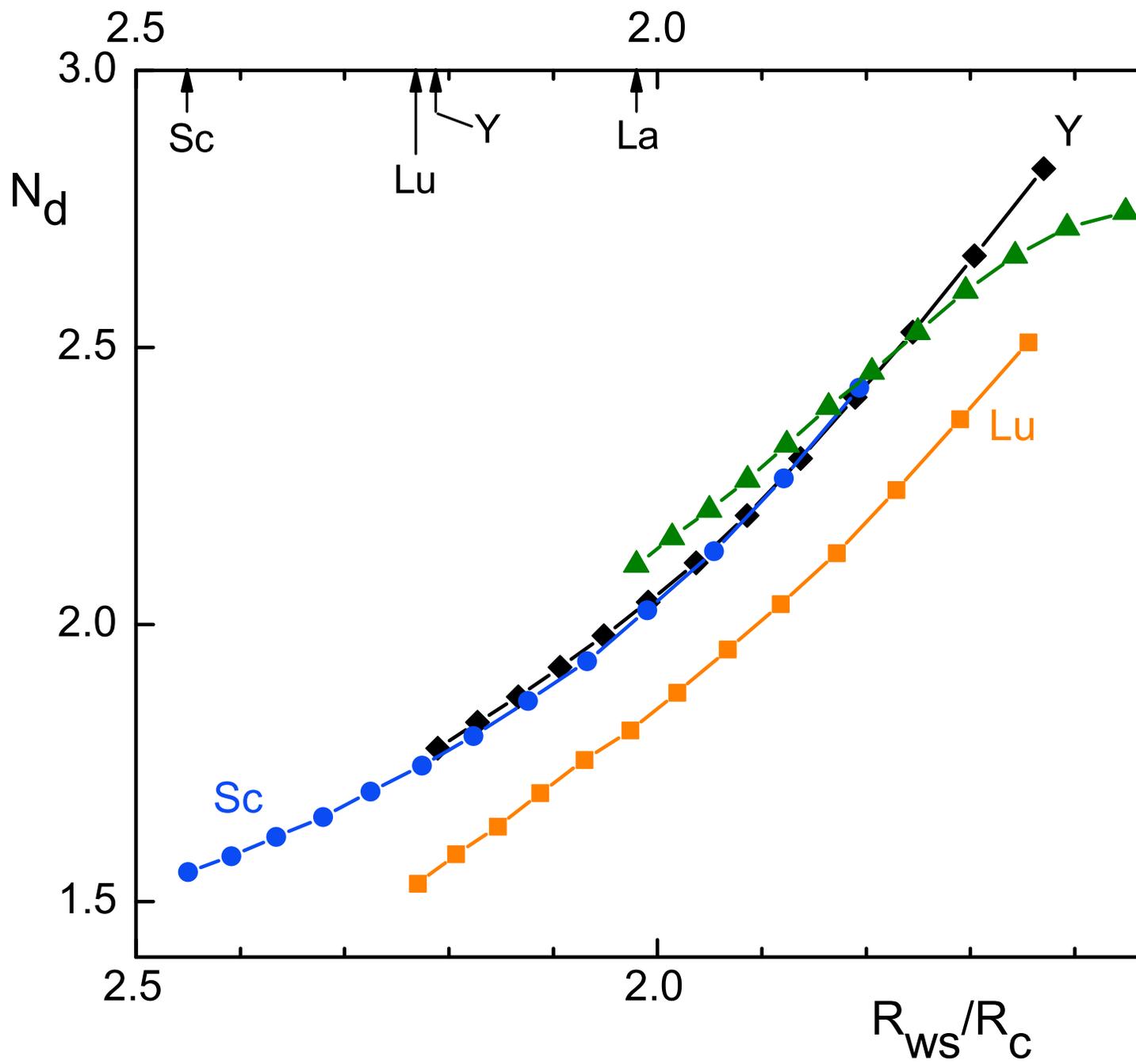
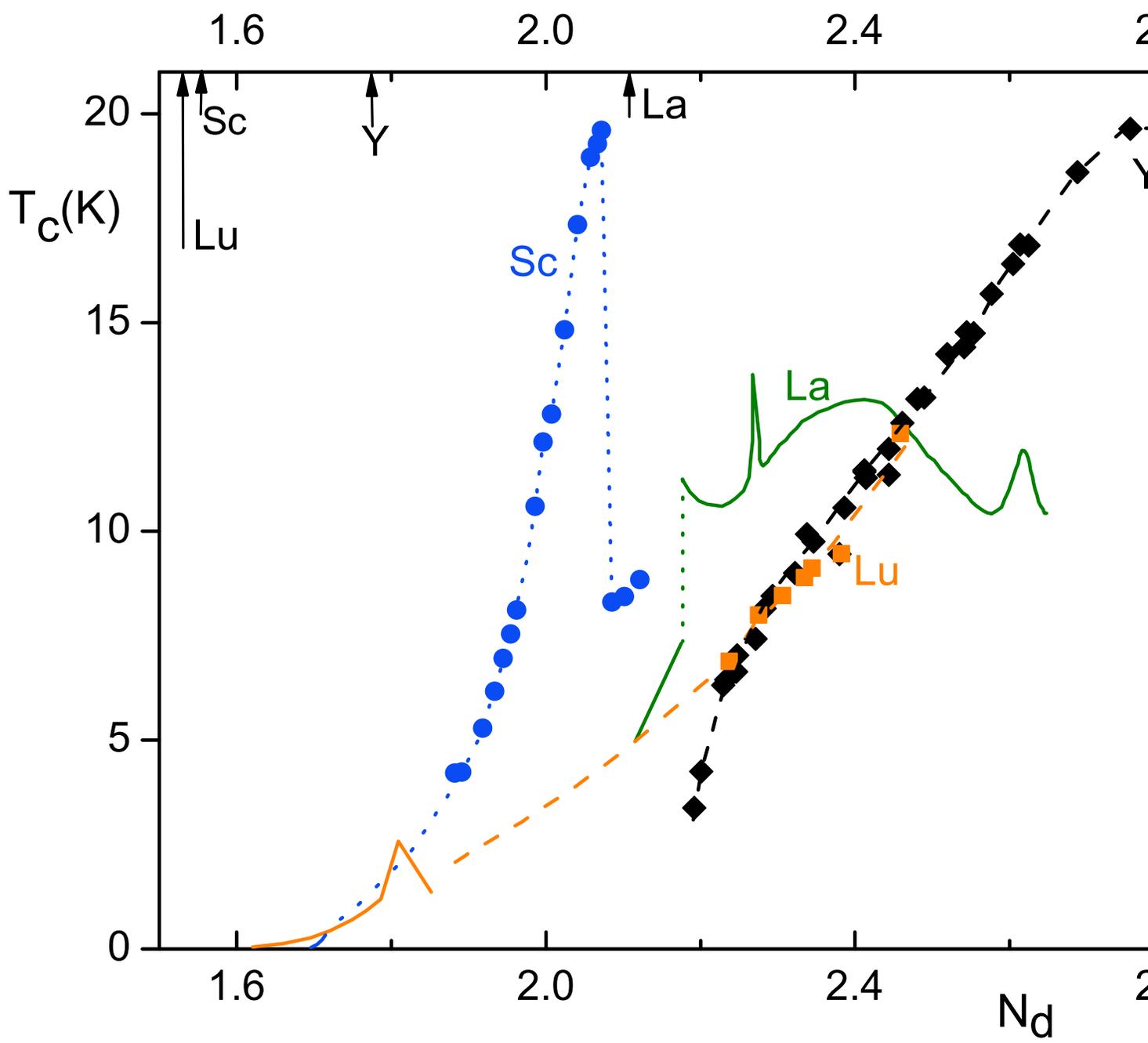
\end{document}